\documentclass[twocolumn,showpacs]{revtex4}

\def\nh4{$\alpha$-(BEDT-TTF)$_2$NH$_4$Hg(SCN)$_4$}

\usepackage{graphicx}
\usepackage{dcolumn}
\usepackage{amsmath}
\begin{document}

\title{Electrodynamics of coupled  
charge-density wave, 2D electron gas systems
}
\author{N.~Harrison
}
\address{
National High Magnetic Field Laboratory, LANL, MS-E536, Los
Alamos, New Mexico 87545
}

\begin{abstract}
The combined inductive and coulombic coupling of an orbitally
quantized two-dimensional electron gas to a one-dimensional
charge-density wave (CDW) is shown to give rise to an anisotropic
quantum fluid in which the Hall electric field, current and gradient
of the CDW phase are all exponentially screened from within the bulk. 
The characteristic penetration depth is similar to the London
penetration depth of relevance to superconductivity.
\end{abstract}

\maketitle 

Two-dimensional electron gas (2DEG) and charge-density wave (CDW)
systems have each been explored extensively for their distinct
electrodynamic properties.  The formation of a Landau gap between the
eigenstates of a 2DEG at high magnetic fields, for example, leads to a
situation in which the diagonal conductivities $\sigma_{xx}$ and
$\sigma_{yy}$ vanish while the off-diagonal Hall components
$\sigma_{xy}=-\sigma_{yx}$ remain finite~\cite{chakraborty1}.  This
leads to dissipationless conduction and also the quantum Hall effect
upon considering interactions.  The formation of a CDW, in contrast,
leads to a gapped state that is electrically insulating, but can
participate in charge transport upon being depinned from the
crystalline lattice by an electric field ${\bf E}$~\cite{gruner1}.

In this paper, I show that the spatial coexistence of these two
phenomena (i.e. a Landau quantized 2DEG and a CDW) occurring on
different sections of Fermi surface, leads to a composite quantum
fluid with electrodynamic properties that are entirely different from
its constituents.  Upon considering the combined inductive and
coulombic coupling between between the two, I obtain a pair of coupled
equations that describe a situation in which the CDW undergoes local
deformations of the phase $\phi$ in response to a divergent electric
field $\nabla\cdot{\bf E}$ and current ${\bf j}$ in the 2DEG. The
anisotropic coupled fluid that results, involving the mutual exchange
of quasiparticles between the 2DEG and states above and below the CDW
gap, exponentially screens variations in ${\bf E}$, ${\bf j}$ and
$\phi$ from within bulk.  The penetration depth $\lambda$, thus
obtained, is similar to that of relevance to
superconductivity~\cite{tinkham1}.  I discuss the limitations of this
effect, and possible experimental configurations in which it may have
been observed.

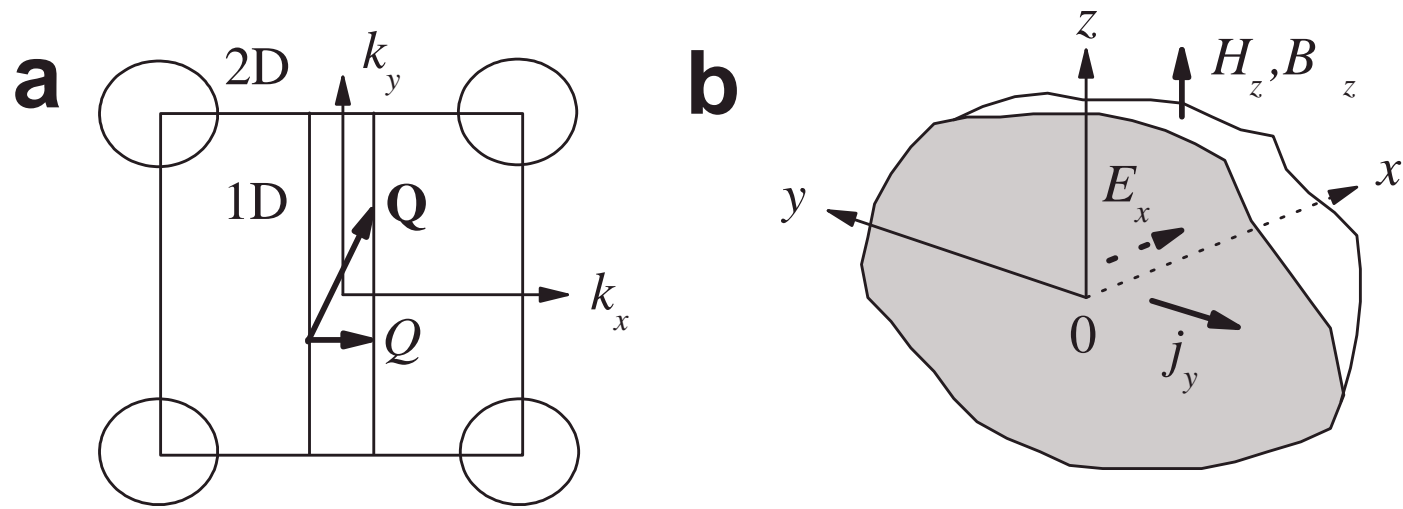
\begin{figure}
\caption{(a) A sketch of the Fermi surface according to the model,
consisting of open and closed sections.  (b) A portion of the surface
of a layered material with this Fermi surface of spatial dimensions
greatly exceeding the penetration depth $\lambda$ (see text).}
\label{diagram}
\end{figure}

To facilitate a comparison with existing
materials~\cite{harrison2,honold1,harrison3,harrison4}, I begin by
considering a bulk anisotropic metal with separate one-dimensional
(1D) electron-like and 2D hole-like quasiparticle dispersions.  On
choosing $\varepsilon_{\rm 1D}=\hbar v_{\rm F}|k_x-k_{\rm
F}|+2t_{b}\cos{k_yb}+2t_{c}\cos{k_zc}$ and $\varepsilon_{\rm
2D}=\varepsilon_{\rm
F}-\hbar^2(|k_x-\frac{\pi}{a}|^2+|k_y-\frac{\pi}{b}|^2)/2m+2t_{c}\cos{k_zc}$
respectively (imposing the limits $t_b\ll\hbar\omega_{\rm c}$ and
$t_c\ll\hbar\omega_{\rm c}$ to simplify subsequent derivations), I
obtain the open and closed sections of Fermi surface depicted in
Fig.~\ref{diagram}(a).  The 1D section is subject to CDW formation
leaving the 2D section intact.

In the nearly-free electron limit, for which the residual short range
Coulomb repulsion is small compared to the electronic bandwidth,
variations of ${\bf j}$ and ${\bf E}$ on the scale of the lattice
parameters $a$, $b$ and $c$ are decoupled from long range variations
of these quatities.  In the present model, I consider variations in
${\bf j}$ and ${\bf E}$ over a characteristic distance $\lambda$
(where $\lambda\gg$~100~nm) that greatly exceeds, not only the lattice
parameters (of order 1~nm), but also the periodicity of the CDW
($2\pi/Q<$~10~nm) and the magnetic length ($l_{\rm m}<$~10~nm at
$\sim$~30~T).  The electrostatic potential whose gradient defines the
long range variations in ${\bf E}({\bf r})=-\nabla V({\bf r})$ can
therefore be considered a long range average that applies over
distances $|\delta{\bf r}|\gtrsim$~10~nm, giving rise to a similar
spatial average in the charge density defined by Poisson's equation
$\nabla^2V({\bf r})=-\rho({\bf r})/\epsilon$~\cite{bleaney1}.

The electrodynamics can be further simplified by considering a
metal-vacuum surface geometry with $x$ normal to the surface at $x=$~0
like that in Fig.~\ref{diagram}(b).  The current $j_y(x)$ flows along
$y$ while the magnetic flux density $B_z(x)$ points along $z$.  In
this way, the spatial variation in all quantities is reduced to an
effective 1D problem in which the $x$-dependence of the electrostatic
potential is given by
\begin{equation}\label{poisson}
    V(x)=\frac{1}{2\epsilon}\int^\infty_0f(x-x^\prime)
    \rho(x^\prime)dx^\prime.
\end{equation}
For the bulk metal described above, $f(x)=|x|$ in contrast to the case
of a single 2DEG for which $f(x)=-\ln|x|$~\cite{macdonald1}.  Each
section of Fermi surface has the potential to yield separate
contributions to the average charge density, enabling the total
density to be expressed as the sum
\begin{equation}\label{charge}
    \rho(x)=\rho_{\rm 1D}(x)+\rho_{\rm 2D}(x).
\end{equation}

The presence of a magnetic field causes holes within the 2D section
(assumed to be spinless) to undergo cyclotron motion, giving rise to a
series of quantized Landau levels of energy $\varepsilon_{\rm
2D}=\varepsilon_{\rm F}-\hbar\omega_{\rm
c}(n+\frac{1}{2})$~\cite{shoenberg1} where $n=$~0, 1, 2,\ldots and
where $\omega_{\rm c}=eB_z/m$ is the cyclotron frequency.  Spatially
varying electric fields modify this cyclotron motion.  Because
$E_x(x)=-\dot{V}(x)$ (the component of ${\bf E}(x)$ along $x$) varies
on a length scale that exceeds the cyclotron radius, this first
derivative of the electrostatic potential with respect to $x$ causes
the centers of the cyclotron orbits to drift at a steady velocity
$v_y(x)=\dot{V}(x)/B_z$~\cite{chakraborty1}; giving rise to the Hall
current.  The second derivative $\dot{E}_x(x)=-\ddot{V}(x)$ modifies
the harmonic oscillation of the holes in the Landau levels by
contributing an additional electrostatic parabolic potetential to
their equation of motion~\cite{macdonald1}.  The cyclotron frequency
thus becomes
\begin{equation}\label{frequency}
    \omega^2(x)=
    \omega^2_{\rm c}-\frac{e\ddot{V}(x)}{m},
\end{equation}
where the `-' sign is appropriate for holes~\cite{macdonald1}. 
Equation (\ref{frequency}) can be verified semiclassically by equating
$md^2{\bf r}^\prime/dt^2$, where ${\bf
r}^\prime\equiv[x^\prime,y^\prime,z^\prime]=[x_0\cos\omega
t,y_0\sin\omega t+v_y(x)t,0]$, to the total force ${\bf
F}=e[B_zv^\prime_y+\dot{V}(x)+\ddot{V}(x)x^\prime,-B_zv^\prime_x,0]$,
where ${\bf v}^\prime\equiv[v^\prime_x,v^\prime_y,v^\prime_z]=d{\bf
r}^\prime/dt$.  Higher derivatives of $V(x)$ perturb the motion
without affecting either $v_y(x)$ or $\omega$.  The change from
$\omega_{\rm c}$ to $\omega$ modifies the eigenenergies of the Landau
levels, so that $\varepsilon_{\rm 2D}=\varepsilon_{\rm
F}-\hbar\omega(n+\frac{1}{2})$.  Because this change is very small, it
is convenient to consider the first derivative
$d\omega/d\ddot{V}(x)=1/2\omega_{\rm c}B_z$ of a Taylor expansion of
$\omega(\ddot{V}$) with respect to $\omega=\omega_{\rm c}$.  This
follows through to an inversely proportionate derivative
$dD(x)/d\ddot{V}(x)=\bar{N}_{\rm 2D}/2\nu\omega_{\rm c}B_z$ in the
Landau level degeneracy $D(x)=\bar{N}_{\rm 2D}\omega/\nu\omega_{\rm
c}$.  Here, $\bar{N}_{\rm 2D}$ is the equilibrium number density of 2D
holes, $\nu=F/B_z$ is the Landau level filling factor and $F$ is the
standard quantum oscillation frequency.  A variation in $D(x)$
therefore corresponds to a variation in charge density
\begin{equation}\label{2Dcharge}
    \rho_{\rm 2D}(x)=-2\alpha\epsilon\ddot{V}(x)+
    \frac{\beta\bar{N}_{\rm 2D}}{\nu\hbar\omega_{\rm c}}\mu^\prime,
\end{equation}
where $|\alpha|\gg$~1 is a dimensionless quantity that depends on the
position of the chemical potential $\mu$ with respect to the highest
occupied Landau level and $\mu^\prime$ accounts for the possibility of
a shift in the chemical potential (this has negligible affect at
integer filling but may be important at half-integer filling).  At
integer Landau level filling factors (when $\mu$ is situated between
two Landau levels), $\alpha_{\rm int}=e\bar{N}_{\rm
2D}/4\epsilon\omega_{\rm c}B_z$~\cite{macdonald1}.  At half-integer
filling factors (when $\mu$ is situated in the middle of a Landau
level), however, $\alpha_{\rm half}=-\infty$.  The latter unphysical
result reflects the fact that the delta function lineshapes of the
Landau levels in an ideal 2D metal cause $\rho_{\rm 2D}(x)$ to jump by
$e\bar{N}_{\rm 2D}/2\nu$ for an infinitesimal change in $\omega$.  The
presence of a finite density of states at $\mu$ introduces a feedback
effect that becomes excessive at half-integer filling factors.  A more
physical result is obtained upon considering a finite relaxation time
$\tau$, which causes the Landau levels to become broadened into
Lorentzians~\cite{shoenberg1}.  In this case, $\alpha=e\bar{N}_{\rm
2D}(1-\beta)/4\epsilon\omega_{\rm c}B_z$, with $\beta_{\rm int}\approx
4/\pi\omega_{\rm c}\tau$ and $\beta_{\rm half}\approx 2\omega_{\rm
c}\tau/\pi$ at integer and half-integer filling factors respectively.

It is clear that in the case of a layered metal without a CDW (in
which case $\rho_{\rm 1D}=$~0), the insertion of Equations
(\ref{2Dcharge}) and (\ref{charge}) into Equation (\ref{poisson})
yields a result only for the uninteresting situation where
$\ddot{V}(x)=0$.  This is one of the reasons why charge tends only to
accumulate beyond the surface in bulk metals, in contrast to single
layer 2DEGs~\cite{macdonald1}.  In order to understand how the
presence of a CDW on the 1D Fermi surface section can change this
situation, it is helpful to consider a Ginzburg-Landau model for CDW
excitations~\cite{gruner1}.  If I ignore high energy amplitude
excitations, derivatives involving time and spin, the local free
energy of a CDW is given by
\begin{equation}\label{DWfree}
    \Phi=\Phi(0)+g_{\rm 1D}\bigg(-\frac{\Delta^2}{2}+\frac{\hbar^2
    v^2_{\rm F}}{4}~\dot{\phi}^2(x)\bigg),
\end{equation}
where $g_{\rm 1D}$ is the density of 1D electronic states, $\Phi(0)$
is the free energy prior to formation of the CDW and $\phi(x)$ is the
local CDW phase.  Here, I define $Q$ as the component of the CDW
nesting vector ${\bf Q}$ parallel to $v_{\rm F}$, while
$Q^\prime(x)=\partial\phi(x)/\partial x\equiv\dot{\phi}(x)$ is the
extent to which it departs from equilibrium.  A change in $Q$ from its
equilibrium value causes the midpoint of the charge gap $2\Delta$ to
shift by an amount $\mu^\prime(x)=\hbar v_{\rm F}k_{\rm
F}Q^\prime(x)/Q=Q^\prime(x)\bar{N}_{\rm 1D}/Qg_{\rm 1D}$ with respect
to $\mu$, where $\bar{N}_{\rm 1D}$ is the mean number density of 1D
carriers~\cite{harrison1}.  This gives rise to a charge
\begin{equation}\label{1Dcharge}
    \rho_{\rm 1D}(x)=-eg_{\rm 1D}\mu^\prime(x);
\end{equation}
not to be confused with the underlying CDW charge modulation that
oscillates with a much shorter periodicity $2\pi/Q\ll l_{\rm c}$,
where $l_{\rm c}=\sqrt{2\hbar F/eB^2_z}$ is the 2DEG cyclotron radius. 
Upon combining Equations (\ref{poisson}), (\ref{charge}),
(\ref{2Dcharge}) and (\ref{1Dcharge}), I obtain $\ddot{V}(x)=-eg_{\rm
1D}\mu^\prime(x)/2\epsilon(\alpha+1)$.  On further making the
approximations $1/\alpha\approx$~0 and $Q\approx 2k_{\rm F}$, this
becomes
\begin{equation}\label{solution}
    \ddot{V}(x)=-\frac{(\eta+\beta) v_{\rm F}B_z}{\nu(1-\beta)}\dot{\phi}(x),
\end{equation}
where $\eta=g_{\rm 1D}/g_{\rm 2D}$ is the ratio of the mean density of
electronic states for the two Fermi surface sections.  By itself,
equation (\ref{solution}) implies that the charge density of the 2D
component of the electronic structure can vary arbitrarily throughout
the bulk because it can always be compensated by a charge density of
opposite sign from the CDW. A small fraction of excess charge
$1/\alpha$ nevertheless remains in order to enable $E_x$ to vary
within the bulk while satisfying Poisson's equation.  This exchange of
charge degrees of freedom is the primary origin of the coupling
between the two subsystems.

Next, I consider the inductive coupling between the 1D and 2D Fermi
surface sections resulting from current flow.  The Hall current
$j_y(x)=(eN_{\rm 2D}/B_z)\dot{V}(x)$ affects the CDW by changing the
local magnetic field $H_z(x)$, where, according to Maxwell's
equations, $j_y(x)=-\partial H_z(x)/\partial x\equiv-\dot{H}_z(x)$. 
It is well established that a change in $H_z(x)$ leads to a change in
the chemical potential $\mu^\prime(x)$ of the
eigenstates~\cite{maniv1}.  These are related via the derivative
\begin{equation}\label{currenteffect}
    \frac{\partial\mu^\prime(x)}{\partial H_z(x)}=\frac{\mu_0\hbar 
    eF}{\gamma mB_z}
\end{equation}
where $B_z=\mu_0(H_z+M_z)$ and where $\gamma_{\rm int}=\eta$ and
$\gamma_{\rm half}=-1$ at integer and half integer filling factors
respectively; the sign change with respect to Reference
\cite{harrison1} accounts for the fact that I consider holes rather
than electrons in the present paper.  Given the relation
$\mu^\prime(x)=\hbar v_{\rm F}\dot{\phi}(x)/2$ identified above, upon
taking its spatial derivative and combining cross derivatives
$\partial\dot{\phi}(x)/\partial
x=\partial\dot{\phi}(x)/\partial\mu^\prime(x)\times\partial\mu^\prime(x)/\partial
H_z(x)\times\partial H_z(x)/\partial x$, I obtain
\begin{equation}\label{solution2}
    \dot{V}(x)=-\frac{\gamma mv_{\rm 
    F}B_z}{2\mu_0\nu e^2\bar{N}_{\rm 2D}}\ddot{\phi}(x).
\end{equation}

Equations (\ref{solution}) and (\ref{solution2}) consitute a pair of
coupled equations that describe the variation of the electric field,
CDW phase gradient and current density with $x$.  On taking the
spatial derivative of Equation (\ref{solution}) in order to combine
them, I obtain
\begin{equation}\label{final}
    \dot{V}(x)=\bigg(\frac{m}{2\mu_0e^2\bar{N}_{\rm 2D}}
    \bigg)\dddot{V}(x)
\end{equation}
where the scalar quantities $\gamma$, $\beta$ and $\eta$ conveniently
cancel on imposing the limit $\omega_{\rm c}\tau\gg$~1.  Equation
(\ref{final}) implies that the Hall electric field, the Hall current
and the phase derivative of the CDW are all exponentially screened
from within bulk so that $E_x(x)=E_{x,0}e^{-x/\lambda}$,
$j_y(x)=j_{y,0}e^{-x/\lambda}$ and
$\dot{\phi}(x)=\dot{\phi}_0e^{-x/\lambda}$.  The characteristic
penetration depth
\begin{equation}\label{lambda}
    \lambda=\sqrt{\frac{m}{2\mu_0e^2\bar{N}_{\rm 2D}}};
\end{equation}
is strikingly similar to the London penetration depth of relevance
to superconductivity~\cite{tinkham1}.

The electrodynamics can be rather succinctly described by a quantum
fluid in which the canonical momentum
\begin{equation}\label{momentum}
    {\bf p}=e{\bf
    A}+\frac{m}{2}\frac{\partial{\bf r}}{\partial t}+
    e({\bf r}\times{\bf B}_0)
\end{equation}
consists of a superfluid-like component coupled to a term
dimensionally equivalent to a Chern-Simons gauge
field~\cite{chakraborty1}.  This comes as no coincidence.  The first
two terms on the right-hand-side of Equation (\ref{momentum})
originate from the existence of a relationship between the current
density and the phase of a wavefunction; a property it shares with
superconductivity~\cite{tinkham1}.  In the present system it is the
CDW phase~\cite{gruner1} rather than the phase of a superconducting
order parameter~\cite{tinkham1} that is of relevance.  The third term
on the right-hand-side of Equation (\ref{momentum}) reflects the fact
that the current in the present system is always encumbered by a
transverse Hall electric field, unlike a supercurrent~\cite{tinkham1}. 
This results from the fact that the Landau levels states in which the
holes reside accommodate the presence of a large steady background
magnetic flux density ${\bf B}_0$ deep within the bulk; i.e
$\lambda\gg x$.  On setting ${\bf p}=$~0, for the geometry considered
in Fig.~\ref{diagram}, this becomes
\begin{equation}\label{vectorpotential}
    {\bf A}=[A_x,A_y,A_z]=[-yB_0,-\mu_0\lambda^2j_y,0].
\end{equation}
It is clear from Equation (\ref{vectorpotential}) that the current
must be non-accelarative (i.e. $\partial j_y/\partial t=0$) in order
to reproduce Equation (\ref{final}), further implying that
$\sigma_{xx}=\sigma_{yy}=$~0.  Thus, upon taking the time derivative
of Equation (\ref{vectorpotential}), I obtain the Hall effect;
\begin{equation}\label{london1}
    E_x=-2\omega_{\rm c}\lambda^2j_y.
\end{equation}
The curl of Equation (\ref{vectorpotential}), in contrast, yields the
London~\cite{tinkham1} equation governing the variation of currents
\begin{equation}\label{london2}
    H_z=-\lambda^2\dot{j}_y+\frac{B_0}{\mu_0},
\end{equation}
albeit in the presence of a background magnetic flux density $B_0$. 
Equation (\ref{final}) can then be obtained upon combining the curl of
Equation (\ref{london2}) with Maxwell's equation, $\dot{H_z}=-j_y$,
and substituting $j_y$ from Equation (\ref{london1}).

There are two ways that $\sigma_{xx}=\sigma_{yy}=$~0 could be achieved
with normal electrons in order for the above electrodynamics to come
into effect.  First, conditions amenable to the realisation of the
quantum Hall effect could be realised at integer filling
factors~\cite{chakraborty1}.  Second, the conductivity can vanish if
the lengthscale over which ${\bf E}$ varies is shorter than the mean
free path $l$.  For example, an electron or hole incident upon the
edge from within the bulk might be accelerated by the electric field
within a distance $\approx\lambda$ only to be decelerated by the same
electric field upon reflection.  However, a more fundamental reason
for the vanishing conductivity may emerge from the physics of the
anisotropic coupled state itself.  Equations (\ref{solution}) and
(\ref{solution2}) imply that as a cyclotron orbit drifts through the
system (in the vicinity of the edge), the inductive and coulombic
coupling causes the CDW to become distorted in its wake.  The charge
of the hole in the cyclotron orbit is therefore effectively screened
by that of the deformed CDW, causing the current-induced anisotropic
coupled state to possess zero net charge.  For this reason, the
coupled state does not respond to longitudinal electric fields giving
rise to $\sigma_{xx}=\sigma_{yy}=$~0, but drifts only orthogonal to a
transverse electric field.  In this way, the resulting electrodynamics
is phenomenologically similar to the quantum Hall effect, except that
here the Hall electric field and the Hall current are confined within
a distance $\lambda$ of the surface, and realisation of
$\sigma_{xx}=\sigma_{yy}=$~0 is not directly dependent on filling
factor $\nu$.  Furthermore, when a transport current flows along the
surface of the material, ones ability to measure the Hall potential
difference will be extremely sensitive to the sample geometry, the
current path and the arrangement of the contacts.  Full realisation of
a dissipationless conducting state requires the penetration depth to
lie within the range $l_{\rm c}<\lambda<d$, where $d$ is mean distance
of CDW pinning sites from the sample surface.  If $\lambda<l_{\rm c}$,
for example, the cyclotron motion will become anharmonic and the
physics will instead be dominated by phenomena associated with edge
states~\cite{chakraborty1}.  If, however, $\lambda>d$, the CDW will be
mostly unable to deform, enabling electric fields and currents to
penetrate into the bulk as in conventional metals.  Realisation of the
inequality $\lambda<l_{\rm c}$ ultimately requires electronic states
of the closed Fermi surface section to vary strongly with
$\varepsilon$, which is most easily realised in the case of a 2D
electronic dispersion in strong magnetic fields.  Realisation of
$\lambda>d$, however, requires the metal to have a high degree purity.

Charge-transfer salts of the type $\alpha$-(BEDT-TTF)$_2M$Hg(SCN)$_4$
may be the most likely candidates to fulfill these criteria.  They
possess 2D and 1D electronic dispersions like that considered in the
model~\cite{mori1}, and the 1D sections are also thought to undergo a
CDW instability at high magnetic fields where the effects of Landau
quantization of the 2D sections are clearly observed~\cite{andres1}. 
These materials are also shown to exhibit a variety of unusual
galvanomagnetic effects.  A pronounced increase in conductivity by as
much as two orders of magnitude is observed at low temperatures in
some crystals, having a behaviour closely resembling that of an
inhomogeneous superconductor~\cite{harrison2}.  This effect is
observed to become weaker at half-integral filling factors.  The Hall
resistance is also measured to exceed the longitudinal
resistance~\cite{honold1}, with a conduction path that is profoundly
sensitive to contact geometry~\cite{harrison3}.  Furthermore,
contactless high frequency measurements yield an inductive response,
part of which can be explained by dissipationless surface
currents~\cite{harrison4}.  A second contribution to the total current
has been shown to arise from the interaction between the composite
CDW-2DEG system and CDW pinning sites in the absence of an electric
field~\cite{harrison1}, although these currents are purely magnetic
${\bf j}_{\rm m}$ in contrast to the electrical ones discussed in the
present model.

In summary, I have shown that inductive and coulombic coupling between
a CDW and 2DEG establishes the existence of a quantum fluid with
significantly different electrodynamics compared to the entities from
which it is composed.  Under ideal conditions, the Hall electric
field, current and gradient of the CDW phase are all exponentially
screened from within the bulk, with a characteristic penetration depth
similar to the London penetration depth.  The present model considers
variation of the electromagnetic fields only with respect to $x$ in
order to simplify the derivation, but the same electrodynamics is
expected to extend to more general geometries.

This work is supported by the Department of Energy, the National
Science Foundation (NSF) and the State of Florida. I would like to 
thank John Singleton for useful comments.

\end{document}